\documentclass[%
 reprint,
superscriptaddress,
longbibliography,
 amsmath,amssymb,
 aps
]{revtex4-2}

\usepackage{graphicx}
\usepackage{dcolumn}
\usepackage{bm}
\usepackage{siunitx}
\usepackage[version=4]{mhchem} 
\usepackage{placeins}
\usepackage{booktabs} 
\usepackage{tabularx,booktabs}
\usepackage{soul}
\usepackage{hyperref}
\hypersetup{
    colorlinks,%
    citecolor=blue,%
    linkcolor=blue,%
    urlcolor=blue
}

\usepackage{lipsum}
\usepackage{siunitx}         
\usepackage{soul}            
\usepackage{wrapfig}
\usepackage[dvipsnames]{xcolor}  
\usepackage{hyperref}
\hypersetup{colorlinks, citecolor=blue, linkcolor=blue, urlcolor=blue}
\usepackage[caption=false]{subfig}
\usepackage{subfig}
\usepackage{ulem}

\newcommand{\orcid}[1]{\href{https://orcid.org/#1}{\includegraphics[width=8pt]{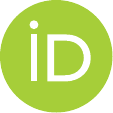}}}

\DeclareSIUnit\rydberg{\text {Ry}}
\begin{document}

\preprint{APS/123-QED}


\title{Controllable Quantum Spin Hall Phases in Bi$_2$Te$_3$-Family van der Waals Heterobilayers}

\author{E. V. C. Lopes \orcid{0000-0002-7981-2161}}
\email{emmanuel.lopes@ilum.cnpem.br}
\affiliation{Ilum School of Science, Brazilian Center for Research in Energy and Materials (CNPEM), Zip Code 13083-970, Campinas, São Paulo, Brazil.}

\author{P. H. Sophia \orcid{0009-0007-5428-0596}}
\affiliation{Ilum School of Science, Brazilian Center for Research in Energy and Materials (CNPEM), Zip Code 13083-970, Campinas, São Paulo, Brazil.}

\author{F. Crasto de Lima\orcid{0000-0002-2937-2620}}
\email{felipe.lima@ilum.cnpem.br}
\affiliation{Ilum School of Science, Brazilian Center for Research in Energy and Materials (CNPEM), Zip Code 13083-970, Campinas, São Paulo, Brazil.}

\author{A. Fazzio\orcid{0000-0001-5384-7676}}
\email{adalberto.fazzio@ilum.cnpem.br}
\affiliation{Ilum School of Science, Brazilian Center for Research in Energy and Materials (CNPEM), Zip Code 13083-970, Campinas, São Paulo, Brazil.}


\begin{abstract}

The tunability and control of topological edge/surface states are crucial for the development of new device applications.
In this work, by combining first-principles calculations and Wannier-based tight-binding methods, we show the emergence of quantum spin Hall phases in van der Waals heterostructures formed by stacking two trivial quintuple layers from the Bi$_2$Te$_3$ family.
We demonstrate the tunability of the edge states under interlayer strain and external electric field effects, suggesting the possibility of switching topological edge states on/off by external control.
Additionally, the quantum spin Hall edge channels remain robust against interlayer twist, highlighting their stability against external perturbations.
Our results provide a new way to create and manipulate two-dimensional topological phases in systems based on Bi$_2$Te$_3$ family, which can be valuable for practical applications, such as topological field effect transistors and spintronic devices. 

\end{abstract}

\maketitle

\section{Introduction}

The family of Bi$_2$Se$_3$ crystals provided a ground-breaking picture in the research of topological phase in three-dimensional materials (3D).
This relevance is supported by the variety of synthesis routes, including chemical vapor deposition~\cite{SRlocatelli2022, APLHasan2012}, molecular beam epitaxy~\cite{AIPADVzeng2013, APLZhang2009, CrystgroWang2021}, and sputtering~\cite{PRMJoaquim2025, JallCompDeng2011}.
It is further reinforced by the possibility of exploring different stoichiometric combinations, such as Bi$_{2-x}$Sb$_x$Se$_{3-y}$Te$_y$~\cite{NatcomunFengliu2022, JPCCWendel2019, PRBAndo2011} as well as heterostructures composed of different materials, including Bi$_2$Se$_3$/Bi$_2$Te$_3$, Sb$_2$Te$_3$/Bi$_2$Te$_3$ and Sb$_2$Se$_3$/Bi$_2$Se$_3$ \cite{NATCOMMeschbach2015, AIPADVzeng2013, SRzhao2013, MTPli2026}.
In this family, the 3D topological phase emerges in films thicker than four quintuple layers (QLs)~\cite{NATPHYSzhang2009, NATPHYSzhang2010,  Kim2011, PRBkobayashi2011}.
A key aspect in the investigation of topological properties in these materials is their robustness against different perturbations, including structural modifications~\cite{NATPHYSliu2014, PRBfocassio2021, ACSCosta2018,AppMaTCapaz2026} and magnetic effects~\cite{FazzioPRB2011, PRBaraujo2024, PRLKim2013, PRBKim2015}.

Their topological character is well established in the bulk limit, but it becomes more subtle in the ultrathin regime, where wave functions localized on opposite surfaces overlap.
This hybridization opens a gap in the Dirac spectrum and, given the time reversal symmetry, can transform the system into either a trivial or a topological 2D insulator \cite{PRLbernevig2006}. Therefore, controlling the coupling between opposite surfaces and the electrostatic asymmetry across the film offers a direct route to manipulate topological phases.
This possibility has been explored using continuum models for thin films of 3D topological insulators \cite{PRBlu2010, PRBli2010, NJPshan2010, PRBmaisel2024}.
In these models, the low-energy bands are described by coupled Dirac states associated with the top and bottom surfaces.
The inter-surface coupling opens a hybridization gap, while an external potential difference or structure inversion asymmetry produces Rashba-like splittings and can drive a gap closing.
Within this minimal picture, the quantum spin Hall phase is commonly expected near the symmetric limit (low external field), whereas sufficiently strong asymmetry tends to suppress the nontrivial phase and drive the system towards a trivial insulating regime \cite{NJPshan2010, PRBli2010}.

Previous continuum models established the essential role of surface-state hybridization and structure inversion asymmetry in ultrathin Bi$_2$Se$_3$-family films.
However, subsequent GW calculations and detailed analysis showed that extracting the topological phase from fitted continuum parameters can be ambiguous, since different parameter sets may reproduce the same dispersion while implying different $Z_2$ indices \cite{PRBforster2015, PRBmaisel2024}.
Therefore, in chemically asymmetric heterobilayers, where band offsets, charge transfer, and finite-k hybridization are important, a direct \textit{ab initio} topological analysis is required.

Realistic heterobilayers, however, do not necessarily follow this symmetric-film scenario.
In a chemically asymmetric stack, the two quintuple layers are inequivalent even at zero applied field.
Band offsets, charge transfer, layer-dependent orbital character, and interface hybridization generate an intrinsic electrostatic asymmetry before any external field is applied.
Moreover, the relevant states may originate from different quintuple layers rather than from the coupled surfaces of a single topological parent compound.
In this case, the band inversion and the minimum gap can evolve away from $\Gamma$ point, requiring an atomistic description of the transition.
As a result, an external field can either enhance or compensate the built-in asymmetry, making it possible to induce, rather than suppress, the topological phase.

In this work, we investigate this mechanism in van der Waals (vdW) X$_2$Te$_3$/Bi$_2$SeTe$_2$ heterostructures, with X = Sb or Bi.
Using first-principles calculations, Wannier-based topological analysis, and tight-binding modeling, we show that the stacking of two topologically trivial quintuple layers can generate a quantum spin Hall phase through interfacial hybridization and charge redistribution.
We demonstrate that the topological character can be controlled by interlayer displacement and by an external electric field, revealing a competition between interface coupling and electrostatic asymmetry.
In contrast to the usual symmetric-film scenario, the applied field can drive the heterobilayer toward the nontrivial regime by compensating the built-in layer asymmetry.
Furthermore, the transition involves band inversion away from the $\Gamma$ point, showing that an atomistic description is essential to capture the topological switching in realistic chemically asymmetric bilayers.

\section{Results and discussion}
\label{sec:Results-discussion}
\subsection{Structural and electronic properties}

\begin{figure*}[htb]
    \centering
    \includegraphics[width=\linewidth, height = 6cm]{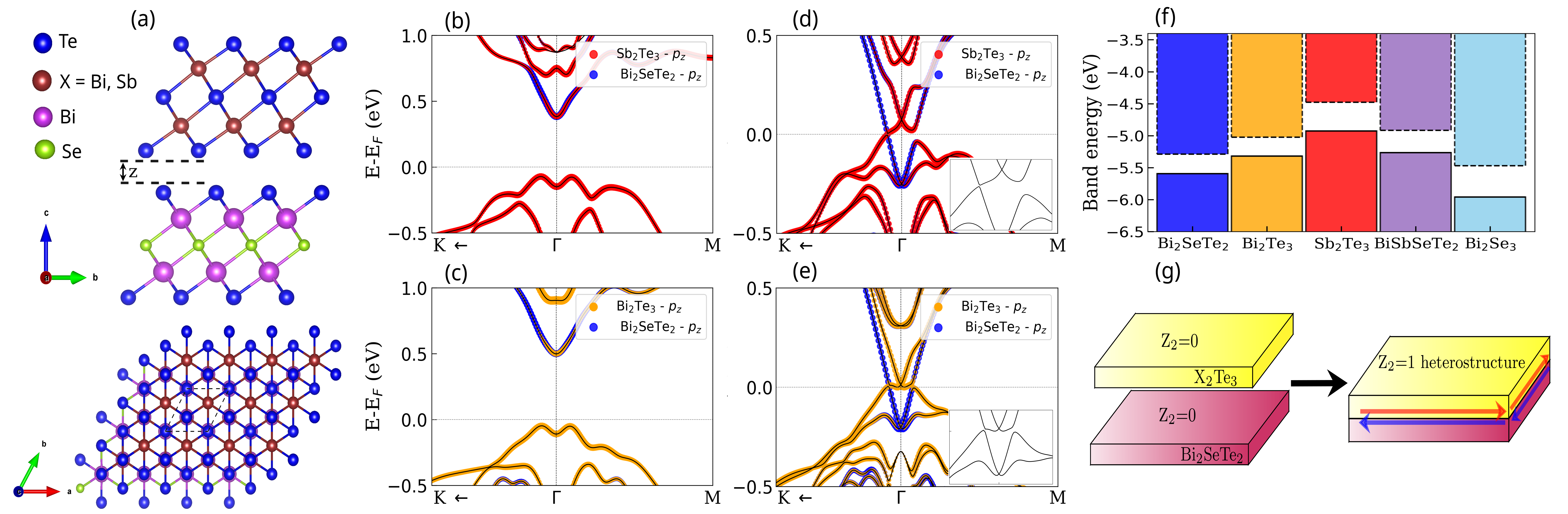}
    \caption{Heterostructure at the top and side views in (a), where the brown spheres represent either Bi or Sb atoms from X$_2$Te$_3$ QL. The contribution of $p_z$ orbitals in the band structure without and with SOC effects is shown in (b) and (d) [(c) and (e)], respectively, for the Sb$_2$Te$_3$/Bi$_2$SeTe$_2$ [Bi$_2$Te$_3$/Bi$_2$SeTe$_2$] heterostructure.
    The insets in (d) and (e) highlight the small band gap in X$_2$Te$_3$ composition.
    Panel (f) shows the schematic band offset relative to the vacuum level for different stoichiometric compositions of the Bi$_2$Te$_3$ QL family.
    The band offset indicates interlayer charge transfer, which enhances spin–orbit coupling and drives a transition from the trivial phase of the isolated QLs to a nontrivial phase in the heterostructure, as schematically shown in (g).}
    \label{fig:painel1}
\end{figure*}

We investigate the heterostructures formed by stacking a single QL of Bi$_2$SeTe$_2$ crystal with a single QL from the same family, X$_2$Te$_3$ (X = Sb, Bi), shown in Fig. \ref{fig:painel1}(a).
Those QLs crystallize in the $P\overline{3}m1$ space group with in plane lattice constants $a= 4.26$\,{\AA} and $4.38$\,{\AA} for Sb$_2$Te$_3$ and Bi$_2$Te$_3$, respectively. 
Such lattice parameters are closer to the Bi$_2$SeTe$_2$, $a =4.29$\,{\AA}, which can lead to a small strain in the vdW interface.
The interface were constructed by combining $1\times1$ supercells of each QL, with the average lattice parameter.
Such choice leads to strain in each QL $\lesssim 1\%$.
Our optimization of the interlayer ($z$) distance shows a $z = 2.89$\,{\AA} and $2.83$\,{\AA} equilibrium separation for Sb$_2$Te$_3$ and Bi$_2$Te$_3$, respectively, stacked on top of Bi$_2$SeTe$_2$, as summarized in Table~[\ref{table}].

\begin{ruledtabular}
\begin{table}[h]
\caption{Structural and electronic properties for the X$_2$Te$_3$/Bi$_2$SeTe$_2$ heterostructure. The considered interface lattice parameter (LP) is given in {\AA}, following the distribute strain in \%, with compressive (tensile) strain indicated by negative (positive) values. The $z$ term represents the equilibrium distance between QLs, in {\AA}, and the charge transfer (CT) in units of e/cm$^2$.}
\begin{tabular}{c c c c c}
  Interface & LP & Strain & z & CT $\times 10^{13}$ \\
  \hline
   Sb$_2$Te$_3$/Bi$_2$SeTe$_2$  & 4.28 & 0.35/-0.35 & 2.89 & 1.10  \\
   Bi$_2$Te$_3$/Bi$_2$SeTe$_2$  & 4.33 & -1.02/1.05 & 2.83 & 0.54  \\
\end{tabular}\label{table}
\end{table}
\end{ruledtabular}

Although the isolated QLs show a centrosymmetric configuration, the heterostructures are intrinsically noncentrosymmetric, giving rise to a Rashba splitting, as can be seen in the band structures without [Fig. \ref{fig:painel1}(b)-(c)] and with [Fig. \ref{fig:painel1}(d)-(e)] spin-orbit coupling (SOC).
Near the Fermi level, only $p_z$ orbitals from each QL layer contribute to the region of valence band maximum and conduction band minimum.
In absence of SOC [Figs. \ref{fig:painel1}(b) and \ref{fig:painel1}(c)], the top of valence band shows the contribution of X$_2$Te$_3$ layer, while the bottom of conduction band exhibits a hybridization of $p_z$ orbitals from both QLs.
The inclusion of SOC reduces the band gap and leads to an inverted band character at $\Gamma$ point [Figs. \ref{fig:painel1}(d) and \ref{fig:painel1}(e)].
This can be seen in the $p_z$ orbital projection from Bi$_2$SeTe$_2$, which, in the presence of SOC, have contribution in the valence band, suggesting a band inversion mechanism driven by layer hybridization and SOC effects.
Additionally, a charge transfer emerges between layers upon stacking, on the order of $0.54 \times  10^{13}$\,e/cm$^{2}$ from Bi$_2$Te$_3$ to Bi$_2$SeTe$_2$ and $1.1\times 10^{13}$\,e/cm$^{2}$ from Sb$_2$Te$_3$ to Bi$_2$SeTe$_2$ QLs, which enhances the relativistic effects emergent from intrinsic inversion symmetry breaking.
This can be seen by the band offset [Fig.~\ref{fig:painel1}(f)], aligned relative to vacuum level, not only for the mentioned systems, but also for other experimentally achievable QLs, Bi$_2$Se$_3$ and BiSbSeTe$_2$~\cite{APLZhang2009, NatcomunPark2015,NatcomunFengliu2022}.
The charge transfer between layers, combined with a noncentrosymmetric nature, induces a potential gradient ($\boldsymbol{\nabla}{V}$), which affects the electronic band structure due to the emergence of a Rashba SOC, given by the Hamiltonian $H_{R}=\lambda_{R}\boldsymbol{\nabla}V\cdot(\boldsymbol{\sigma}\times{\bf{k}})$.
Here, $\lambda_{R}$ denotes the Rashba SOC strength, $\boldsymbol{\sigma}$ the Pauli matrices, and $\boldsymbol{k}$ the momentum coordinates. 

This microscopic picture differs from the conventional thin-film limit of a three-dimensional topological insulator.
In symmetric ultrathin films, the low-energy bands are usually described in terms of hybridized top and bottom surface states of the same parent compound.
In the present heterostructures, however, the states near the Fermi level arise from chemically distinct quintuple layers.
The band offsets, interfacial charge transfer, and hybridization between layer-resolved $p_z$ orbitals generate an intrinsic electrostatic asymmetry already at zero external field.
Therefore, the band ordering is not controlled only by the thickness-dependent coupling between equivalent surfaces, but by the relative alignment and overlap of orbitals across the interface.

\subsection{Topological properties}

\begin{figure*}[htb]
    \centering
    \includegraphics[width=\linewidth, height = 7cm]{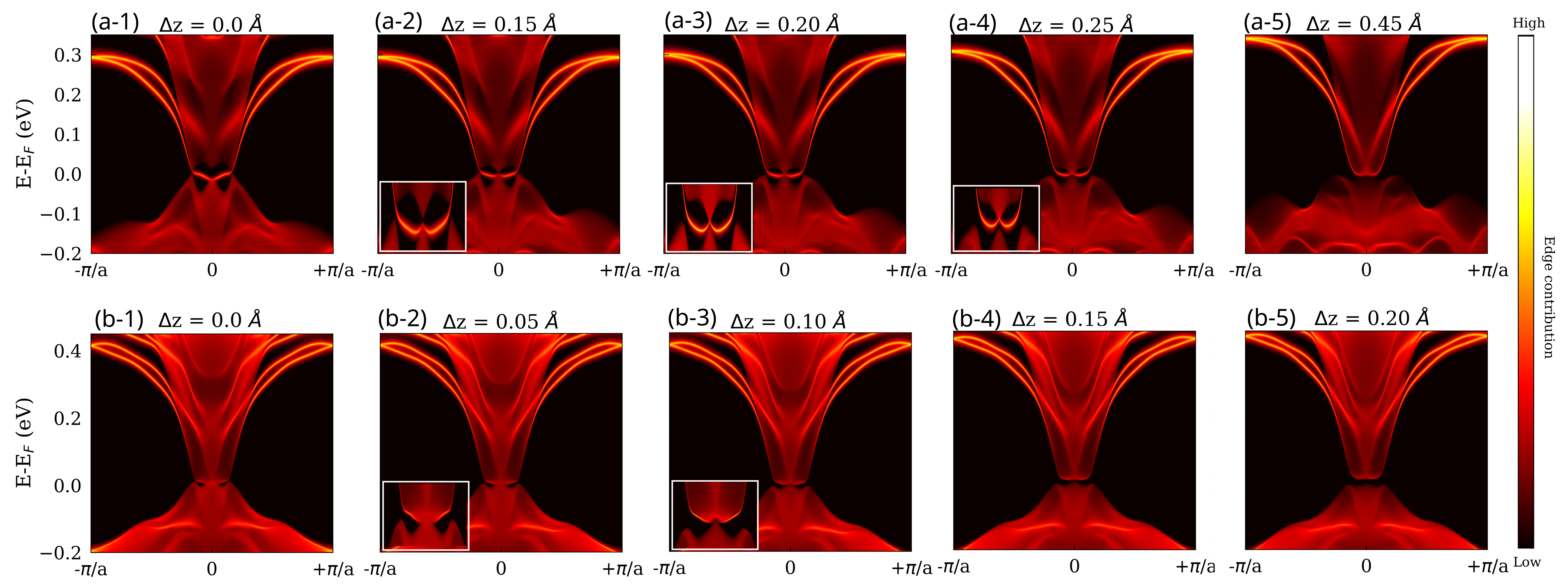}
    \caption{Evolution of topological edge states from Sb$_2$Te$_3$/Bi$_2$SeTe$_2$ (a) and Bi$_2$Te$_3$/Bi$_2$SeTe$_2$ (b) heterostructures under different interlayer separation distances starting from the equilibrium ($\Delta{z}=0.0$ {\AA}). The insets in (a-2)-(a-4) and (b-2)-(b-3) highlight the region of edge states emergence, indicating either a topological [(a-2) and (b-2)] or trivial [(a-3), (a-4) and (b-3)] behavior.}
    \label{fig:Evolution}
\end{figure*}

The inversion of the layer orbital contribution upon inclusion of SOC suggests a nontrivial topology, emergent from the interplay between layer stacking and fully relativistic effects.
We confirm the nontrivial behavior by calculating the Z$_2$ invariant through the Wannier charge center method, where both heterostructures show Z$_{2} = 1$, while their isolated single QL counterparts show Z$_{2} = 0$, as schematically shown in Fig.~\ref{fig:painel1}(g).
In addition to the single QL analysis, in Fig. S1 of the Supplemental Material \cite{supplemental}, we show that the stacking of two QLs with the same composition also exhibits a topologically trivial character. This reinforces the crucial role of electrostatic asymmetry in driving the system into a nontrivial regime, also discussed in Fig. S2.
It is important to note that the other QLs from Fig.~\ref{fig:painel1}(f) in a heterostructure with Bi$_2$SeTe$_2$ also show a nontrivial phase, as can be seen in Fig. S3.
The corresponding topological edge states are shown in Figs. \ref{fig:Evolution}(a-1) for Sb$_2$Te$_3$/Bi$_2$SeTe$_2$ and \ref{fig:Evolution}(b-1) for Bi$_2$Te$_3$/Bi$_2$SeTe$_2$.
Unlike the stacking of more than four X$_2$Te$_3$ QLs, which shows a 3D topological insulator phase \cite{Xia2009, Kim2011}, in these heterostructures the 2D quantum spin Hall nontrivial phase already emerges at the two QL limit.

Both isolated QLs display a trivial character, therefore the increase in layer separation in the heterostructure ($\Delta z \rightarrow \infty$) may restore the trivial phase.
Figure \ref{fig:Evolution} shows the edge state behavior under different $\Delta{z}$ displacements for X$_2$Te$_3$/Bi$_2$SeTe$_2$ heterostructures, where panels (a) and (b) correspond to the Sb- and Bi-based compositions, respectively.
Here, $\Delta{z}=0$ denotes the optimized equilibrium distance between the QLs, which shows topological crossings in ${\pm}\pi/a$ connecting the valence to conduction bands [Figs.~\ref{fig:Evolution}(a-1) and ~\ref{fig:Evolution}(b-1)].
It can be noted that, by increasing the interlayer distances, a reduction in the overlap between $p_z$ orbitals from distinct QLs leads to a non-inverted bulk band structure, therefore affecting the emergence of the Dirac crossing in the valence band.
In Sb$_2$Te$_3$/Bi$_2$SeTe$_2$, for interlayer separations up to $\Delta{z}= 0.15$ {\AA}, the system exhibits a nontrivial topological character, with the edge states emerging from the valence band and connecting to the conduction band, as highlighted in the inset of Fig. \ref{fig:Evolution}(a-2).
For larger displacements, $\Delta{z}>0.15$ {\AA}, the competition between intrinsic SOC and Rashba effect is reduced to the point where the interfacial interaction no longer induces band inversion.
In contrast, the Bi$_2$Te$_3$/Bi$_2$SeTe$_2$ heterostructure is less robust against interlayer displacement.
This behavior arises from the smaller charge transfer between the QLs in this system, that is, smaller Rashba term.
The system remains topological up to $\Delta{z}=0.05$ {\AA} [Fig. \ref{fig:Evolution}(b-2)], while at $\Delta{z}=0.10$ {\AA} it becomes trivial [Fig. \ref{fig:Evolution}(b-3)].
As shown in Figs. \ref{fig:Evolution}(a-4)-(a-5) and \ref{fig:Evolution}(b-4)-(b-5), the trivial band gap increases with further vertical separation.
In contrast, the reduction in interlayer distance leads to the increase in topological gap in bulk, widening the energetic region of quantized transport in X$_2$Te$_3$/Bi$_2$SeTe$_2$ heterostructure [Fig.~S4]. 

\begin{figure}[htb]
    \centering
    \includegraphics[width=\columnwidth, height = 5.5cm]{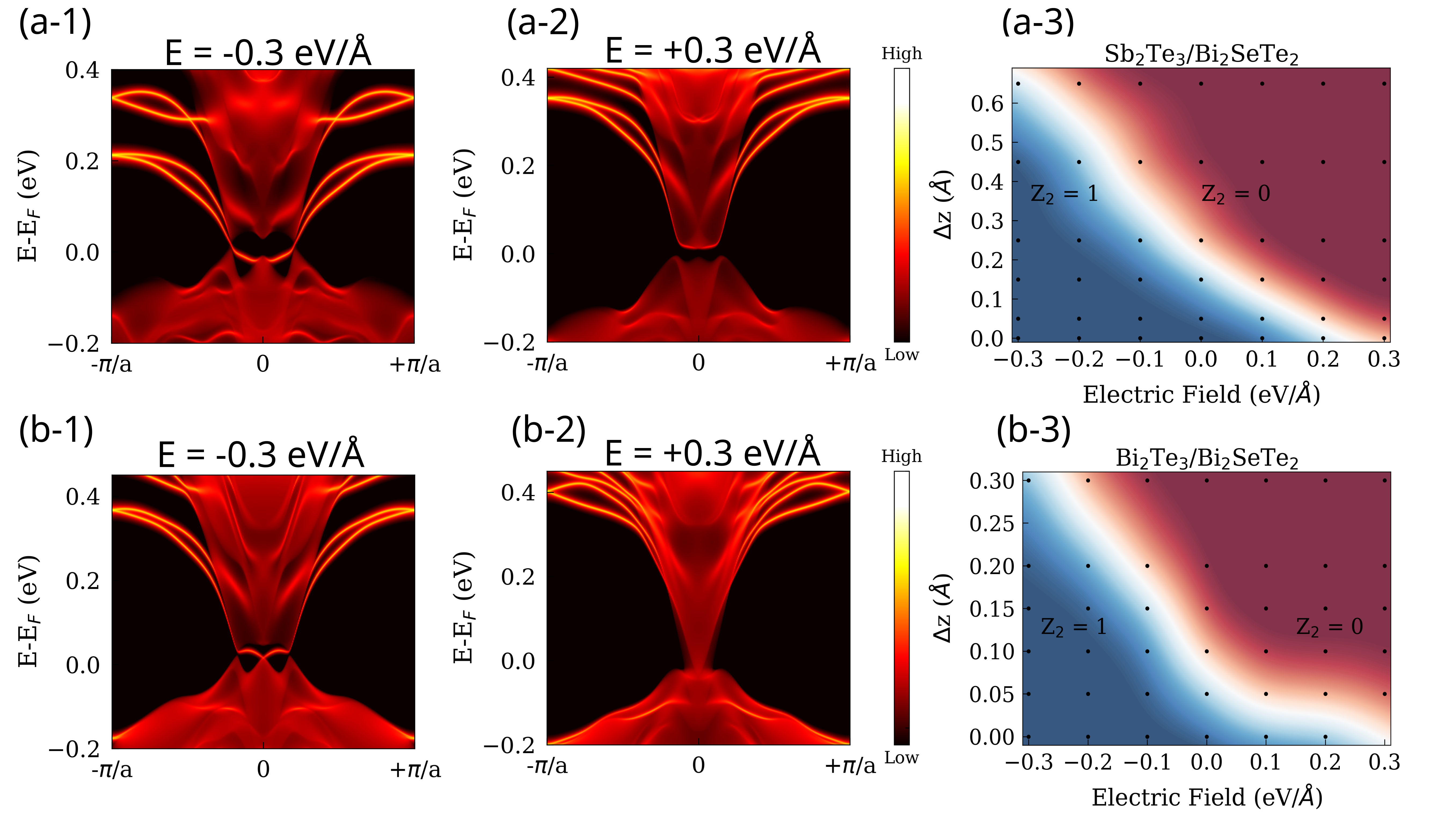}
    \caption{Switchable quantum spin Hall edge states under the influence of an external electric field applied along directions that enhance [(a-1) and (b-1)] and suppress [(a-2) and (b-2)] the topological properties of Sb$_2$Te$_3$/Bi$_2$SeTe$_2$ (a) and Bi$_2$Te$_3$/Bi$_2$SeTe$_2$ (b) at the vertical equilibrium distance. The figures (a-3) and (b-3) display the topological phase diagram which takes into account the electric field, displacement and topology. The red map indicates the trivial regime, whereas the blue indicates the nontrivial. }
    \label{fig:Efield}
\end{figure}

Positive or negative interlayer displacement leads to the suppression or enhancement of the topological band gap due to hybridization and interfacial charge transfer effects \cite{PRBlima2017, Bordoloi2DMat2025}, which suggests that external electric fields can play a significant role in these systems.
For both heterostructures, at their equilibrium separation, an external electric field enables a switchable topological phase depending on its orientation.
This switchable behavior is observed in both heterostructures under an external electric field normal to the surface.
For instance, ${E} =-0.3$\,eV/{\AA}, favoring the increase in charge transfer, the topological bulk band gap at $\Gamma$ point increases [Figs. \ref{fig:Efield}(a-1) and \ref{fig:Efield}(b-1)].
By applying the external electric field in the opposite way [Figs. \ref{fig:Efield}(a-2) and \ref{fig:Efield}(b-2)], a trivial gap appears for Sb$_2$Te$_3$, with the topological edge states connecting conduction and valence bands vanishing, while Bi$_2$Te$_3$ reaches a critical point of phase transition. 
It is worth noticing that higher fields lead to a trivial phase.
This switchable quantum spin Hall edge channels, observed in both systems under external field effects, makes X$_2$Te$_3$/Bi$_2$SeTe$_2$ promising candidates for topological field effect transistors \cite{Qian2014, Vandenberghe2017, SinghReview2026, NLFazzio2015}.
As previously shown, interlayer displacement affects the nontrivial phase of the heterostructures.
These effects compete with the influence of external electric fields, thereby also affecting the topological phase.
Figures \ref{fig:Efield}(a-3) and \ref{fig:Efield}(b-3) show the topological phase diagram as a function of interlayer displacement and external electric field for Sb$_2$Te$_3$- and Bi$_2$Te$_3$-based heterostructures, respectively.
At the equilibrium separation, the topological behavior of the Sb$_2$Te$_3$/Bi$_2$SeTe$_2$ heterostructure remains resilient for electric field at the order near to $+0.2$ eV/{\AA}.
However, for interlayers distance on the order of $\Delta{z}=+0.15$ {\AA}, only negative or zero external fields displays a non-zero topological index, illustrating the competition mechanism.
While in Sb-based heterostructure the nontrivial topological phase can be achieved with moderate electric fields for separations up to $+0.6$ {\AA}, the Bi-based heterostructure, due to its reduced charge transfer, shows weaker tunability under external electric fields.

The phase diagrams in Figs.~\ref{fig:Efield}(a-3) and \ref{fig:Efield}(b-3) summarize the competition between interfacial hybridization and electrostatic asymmetry.
Increasing the interlayer distance reduces the overlap between the $p_z$ orbitals of the two QLs and drives the system toward the trivial regime.
Conversely, an external electric field changes the relative alignment of the layer-resolved states.
Depending on its direction, the field can either enhance the intrinsic asymmetry of the heterobilayer or compensate it, restoring the band inversion and inducing the quantum spin Hall phase.
This explains why the field response differs from the symmetric thin-film scenario, where increasing structure inversion asymmetry usually suppresses the nontrivial phase.
An independent atomistic Slater–Koster model, discussed in the Supplementary Material, reproduces the same qualitative phase switching [Fig. S5], confirming that the mechanism does not depend on the details of the Wannier construction.
It is worth noting that the critical point of the topological phase transition can change depending on the interlayer distance obtained using different vdW corrections and exchange-correlation approaches~\cite{ZungerPRB2011, LuoPRB2012, NechaevPRB2013, ChengPRB2014}, nevertheless, the topological phase can still be tuned through interlayer displacement and/or an external electric field.
To assess this dependence explicitly, we also performed Heyd-Scuseria-Ernzerhof screened hybrid functional (HSE06) calculations for the Sb$_2$Te$_3$/Bi$_2$SeTe$_2$ heterostructure. At the equilibrium separation obtained with vdW-optB86b, HSE06 predicts a trivial phase; however, reducing the interlayer separation drives the system into a quantum spin Hall phase, reproducing the same qualitative evolution obtained with PBE. Moreover, at the PBE-D3 functional \cite{Grimme2010} equilibrium separation of $2.69$\,{\AA}, which is closer to experimentally reported inter-QL distances for the Bi$_2$Te$_3$ family~\cite{NakajimaJPCS1963}, both PBE and HSE06 predict a non-trivial phase (see Supplemental Material section V). Therefore, although the gap and critical separation depend on the computational approximation, the underlying competition between inter-QL hybridization and electrostatic asymmetry remains robust.

\subsection{Twisted heterostructure}

\begin{figure}[htb]
    \centering
    \includegraphics[width=\columnwidth, height = 7.5
    cm]{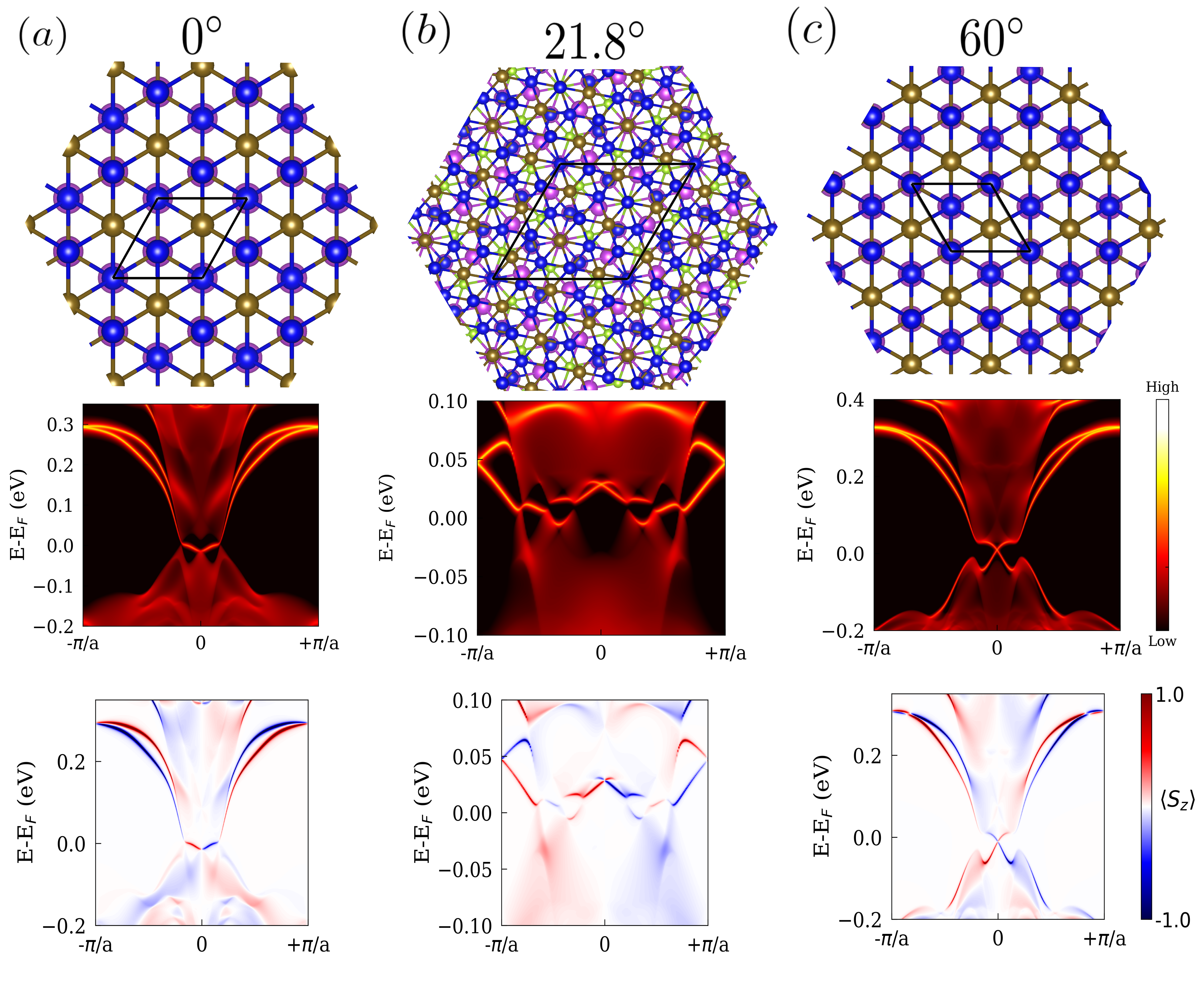}
    \caption{Sb$_2$Te$_3$/Bi$_2$SeTe$_2$ structure (upper figures) and their corresponding edge states band structure (middle figures) and spin projection (lower figures) under different rotation angles, (a) for $0^{\circ}$, (b) for $21.8^{\circ}$, and (c) for $60^{\circ}$.}
    \label{fig:Twist}
\end{figure}

As previously discussed, interlayer displacements highly affects the emergence of a nontrivial phase in both heterostructures.
This naturally raises the question of whether alternative staking configurations or interlayer twisting can also affect the topological properties of the heterostructures.
Motivated by this, we investigate the topological robustness of the Sb$_2$Te$_3$/Bi$_2$SeTe$_2$ heterostructure under different twist configurations.
In addition to the $0^{\circ}$ [Fig.~\ref{fig:Twist}(a)], we consider interlayer twists of $21.8^{\circ}$ [Fig.~\ref{fig:Twist}(b)], which give rise to a Moiré pattern, and for $60^{\circ}$ [Fig.~\ref{fig:Twist}(c)].
Our results show that the topological behavior remains unchanged due to rotation.
Due to the interlayer twist, the K point of Sb$_2$Te$_3$ shifts to a different position in the Brillouin zone relative to that of Bi$_2$SeTe$_2$ counterpart, which affects the band dispersion, although does not affect the topological behavior.
For a $60^{\circ}$ of rotation, which has a total energy $1.9$ meV/atom higher than the $0^{\circ}$ configuration (the most stable configuration), the crossing at $\Gamma$ point becomes more clearly visible due to a slightly larger band gap and reduced bulk states hybridization.
The energy difference between $0^{\circ}$ and $60^{\circ}$ in Sb$_2$Te$_3$-based heterostructure lies on the same order of other 2D vdW systems~\cite{FazzioPRB2018, PRLMarom2010, JPCCLiu2012}.
It is important to highlight that other stacking types, although less energetically favorable, also show the topological phase. 

\section{Conclusions}

In summary, our results demonstrate the emergence of a quantum spin Hall phase in heterostructures composed of two trivial distinct quintuple layers from the Bi$_2$Te$_3$ family, the X$_2$Te$_3$ (X = Sb, Bi) in a van der Waals interaction with Bi$_2$SeTe$_2$.
By means of first-principles calculations and tight-binding methods, we show that the layer stacking leads to charge transfer which, combined with the intrinsic noncentrosymmetric character, drives a Rashba effect playing a crucial role in the emergence of a nontrivial phase.
This mechanism is further confirmed by the investigating the interlayer displacement, where, by increasing the separation, a trivial phase emerges in both heterostructures.
In contrast, reducing the distance increases the topological band gap and enhances the visibility of the Dirac crossing in Sb$_2$Te$_3$-based system.
Additionally, our investigation includes a mechanism to control the quantum spin Hall edge channels by switching it on and off under an external electric field, suggesting potential applications in topological field-effect transistors.
Our results not only show a route to induce topological properties in dimensionally reduced systems from the Bi$_2$Te$_3$ family, well known for exhibiting nontrivial phases primarily in three-dimensional scenarios, but also elucidate their robustness against external perturbations, such as electric field, interlayer displacement, and interlayer twist.
Due to their switchable topological properties under an external electric field and high experimental controllability, the X$_2$Te$_3$/Bi$_2$SeTe$_2$ heterostructures are promising candidates for applications in topological field-effect transistors and devices that require nontrivial systems with two-dimensional architecture.

\section*{Methodology}

First-principles calculations were performed within density functional theory (DFT) employed in Vienna \textit{ab initio} simulation package (VASP) \cite{PRBkresse1996}.
The electronic density was described by means of projector-augmented-wave pseudopotentials using the generalized gradient approximation (GGA) in the Perdew-Burke-Ernzehof formalism~\cite{PRLperdew1996}.
The cutoff in kinetic energy was set at 300 eV. 
For the non twisted unit cells, the Brillouin zone were sampled using the $11 \times 11 \times 1$ Monkhost-Pack grid \cite{PRBmonkhorst1976}, while for the twisted we mapped using $5 \times 5 \times 1$ grid.
To avoid any interaction with periodic images, we set vacuum distance between layers as 15 {\AA}.
Structural relaxations were performed with fixed in-plane lattice vectors.
For the untwisted heterostructures, the in-plane lattice parameter was fixed at the average of the optimized lattice parameters of the two isolated QLs, resulting in strains below $1\%$ in each layer.
For the twisted heterostructure, the lattice vectors and twist angle of the commensurate supercell were kept fixed.
In all cases, all internal atomic coordinates, including those within the individual QLs, were fully relaxed.
The atomic positions were relaxed until the forces were below the 0.01 eV/{\AA}, with a total energy convergence criterion of 10$^{-6}$\,eV.
We employed the non-local van der Waals functional optB86b~\cite{PRBKlime2011} to describe the interlayer van der Waals interaction.
Unless otherwise mentioned, all calculations were conducted with the inclusion of spin-orbit coupling (SOC). 
External electric field was included perpendicularly to the heterostructures plane in the saw-tooth potential description and the charge transfer were investigated by means of Bader charge method \cite{Tang2009}.
We generate the twisted Sb$_2$Te$_3$/Bi$_2$SeTe$_2$ structure within the SAMBA package facility~\cite{Araujo2025}, where, under the criteria up to 100 atoms in unit cell, only the twisted angle of $21.8^{\circ}$ gives a commensurable cell.
We extracted the tight-binding model from first-principles results with Wannier90 package \cite{Wannier90}. The maximally localized Wannier functions were constructed using \textit{s} and \textit{p} orbitals of all atoms in the heterostructures, resulting in a total of 80 Wannier bands. The disentanglement procedure was performed within an energy window ranging from -12.9 eV to 6.9 eV, builded on the same Monkhost-Pack grid. We calculate the topological properties, such as Z$_2$ invariant and Topological states within the Wanniertools code~\cite{WannierTools}. 

\section*{Data availability statement}

All data that support the findings of this study are included within the article (and any supplementary files).

\begin{acknowledgments}
The authors acknowledge financial support from the Brazilian agencies FAPESP (grants 23/09820-2, 24/00989-7, 2024/22048-0, and 25/13106-9), CNPq, INCT-Nanocarbono, INCT-Materials Informatics, and the LNCC (SDumont, Projects DIDMat2 and SCAFMat2) for computer time.

\end{acknowledgments}

\bibliography{bib}%

\end{document}